\documentclass[times,a4paper]{article}
\pdfoutput=1

\usepackage{cite}
\usepackage{times}
\usepackage{balance}
\usepackage{subfig}
\usepackage[utf8]{inputenc}
\usepackage{amsmath}
\usepackage{amsfonts, amssymb}
\usepackage{listings}
\usepackage{graphicx}
\usepackage{marvosym}
\usepackage{comment}
\usepackage{textcomp}
\usepackage{enumitem}
\usepackage{color}
\usepackage{array}
\usepackage[algo2e,linesnumbered,ruled,vlined]{algorithm2e}
\usepackage{authblk}
\usepackage{fullpage}
\usepackage[colorlinks,bookmarksopen,bookmarksnumbered,citecolor=blue,urlcolor=red]{hyperref} 

\DeclareCaptionType{copyrightbox}

\newcommand{\hide}[1]{}

\newcommand{\highlight}[1]{{\color{red}\textbf{#1}}}
\renewcommand{\highlight}[1]{#1}

\providecommand{\lstparallel}[1]{\mathbf{parallel_{#1}}}
\providecommand{\KNNGPU}{\textsf{K-NN$_{\text{GPU}}$}}
\providecommand{\KNNGPUCACHE}{\textsf{K-NN$_{\text{GPU}}^{CACHE}$}}
\providecommand{\KNNGPUCOALESCE}{\textsf{K-NN$_{\text{GPU}}^{COALESCE}$}}
\providecommand{\KNNCPU}{\textsf{K-NN$_{\text{CPU}}$}}
\providecommand{\KNNBASELINE}{\textsf{K-NN$_{\text{BASELINE}}$}}

\begin{document}

\newtheorem{lemma}{Lemma}
\newtheorem{property}{Property}
\newtheorem{definition}{Definition}

\title{Manycore processing of repeated k-NN queries\\over massive moving objects observations}
\author{Francesco Lettich \thanks{lettich@dais.unive.it}}
\author{Salvatore Orlando \thanks{orlando@unive.it}}
\author{Claudio Silvestri \thanks{silvestri@unive.it}}
\affil{Dipartimento di Scienze Ambientali, Informatica e Statistica, Università Ca' Foscari\\Via Torino 155, Venice, Italy}

\maketitle

\begin{abstract}
The ability to timely process significant amounts of continuously updated spatial data is mandatory for an increasing number of applications.
\hide
{
Parallelism enables such applications to face this data-intensive challenge and allows the devised systems to feature low latency and high scalability. 
In this context, commodity graphics processing units (GPUs) represent
extremely interesting low-cost and high-efficient parallel architectures, even though data-intensive applications pose serious challenges to GPUs due to the massive input/output data streams exchanged and workload distribution issues.  
}
In this paper we focus on a specific data-intensive problem concerning the repeated processing of huge amounts of \emph{k} nearest neighbours (k-NN) queries over massive sets of moving objects, where the spatial extents of queries and the position of objects are continuously modified over time. In particular, we propose a novel hybrid CPU/GPU pipeline that significantly accelerate query processing thanks to a combination of ad-hoc data structures and non-trivial memory access patterns. 

%
\hide
{
We devise a novel hybrid CPU/GPU pipeline to significantly accelerate query processing. 
The devised system relies on an ad-hoc \emph{point-region quadtree} based spatial index coupled with an iterative approach for batch-based query processing, aiming to minimize the amount of computations per query while producing workloads suitable for GPU processing. 

In particular, an iterative approach is needed due to the adoption of a tree-based index and to the fact that k-NN queries spatial extent is not known beforehand. This leads to a problem decomposition that results in a set of independent data-parallel tasks, suitable for GPU processing, produced on the fly as query computation progresses.
The adoption of such index allows also to tackle effectively a broad range of spatial object distributions, even those very skewed. 

Also, to deal with the architectural peculiarities and limitations of GPUs 
we adopt non-trivial memory access patterns that avoid the need of locked memory accesses and favor cached or coalesced memory accesses, thus enhancing the overall memory throughput.
}

To the best of our knowledge this is the first work that exploits GPUs to efficiently solve repeated k-NN queries over massive sets of continuously moving objects, even characterized by highly skewed spatial distributions. 
In comparison with state-of-the-art sequential CPU-based implementations, our method highlights significant speedups in the order of 10x-20x, depending on the datasets, even when considering cheap GPUs.

\end{abstract}


\section{Introduction}
\label{sec:intro}

An increasing amount of applications need to process massive
spatial workloads. Specifically, we consider applications in settings
where spatial data is continuously produced over time and needs to be processed
rapidly, e.g., scenarios involving services or applications mainly tailored for mobile device infrastructures - such as \emph{Location-Based Services} (LBS) or \emph{Location-Based Social Networking} applications (LBSN) \cite{pelekis2014mobility}, and others, like behavioural simulations or Massively Multiplayer Online Games (MMOG) \cite{sowell2013experimental}.  
In these applications, very large populations of continuously \emph{moving
objects} frequently update their positions and issue some kind of query to 
look for other objects within their interaction area. The resulting massive workloads pose
new challenges to data management techniques. 
In this context we consider k-NN queries, i.e., globular queries whose spatial extent is dictated by the dislocation of the \emph{k} nearest objects with respect to their centers.

To enable parallel processing and optimizations, and thus manage
the targeted workloads in a scalable manner, we recur to an approach based on time discretization.
In this sense, we partition the time in intervals (or \textit{ticks}), assign location updates and queries to the ticks in
which they occur and process the updates and queries in the resulting
batches such that query results are reported after the end of each tick.
This approach has the effect of replacing the processing of a large
number of independent and asynchronous queries with the \emph{iterated processing
of spatial joins} between the last known positions of all the moving objects at the end of a
tick and the queries issued during the tick. In other words we trade (slightly) delayed processing of queries for increased throughput, and therefore care is needed to ensure acceptable delays.
To achieve high performance and scalability we also exploit a platform encompassing an off-the-shelf general-purpose microprocessor (CPU) coupled with a
\textit{Graphics Processing Unit} (GPU) that features hundreds of 
processing cores. To benefit from GPUs exploitation,
limitations and peculiarities of these architectures must be carefully taken into
account \cite{lee2010debunking}. Specifically, individual GPU cores are slower than those
of a typical CPU, while some memory access patterns may cause serious performance degradation due to
contention and serialization of memory accesses. Effective query processing techniques are needed to address
these limitations: multiple cores must work together to efficiently process queries in parallel for most of the time, and must coordinate their activities to ensure high memory bandwidth.

Our contributions can be summarized as follows: first, we introduce a framework for repeated processing of massive k-NN queries over massive moving object observations. Second, on the basis of this framework we introduce \KNNGPU, an hybrid CPU-GPU k-NN query processing pipeline able to efficiently compute batches of k-NN queries. \KNNGPU\ processes queries using an iterative approach, where at each iteration 
blocks of close objects are scanned in parallel to update the result sets of blocks of k-NN queries. 
At each iteration we exploit proper memory access patterns coupled with ad-hoc memory layouts of objects and queries, which are arranged according to some spatially preserving function: this entails workloads suitable for GPU processing and improves memory bandwidth due to better exploitation of GPUs caching and coalescing capabilities. Finally, we carry out an extensive set of experiments to study how \KNNGPU\ varies its performance for different datasets and algorithm parameters. 
We also compare \KNNGPU\ with a GPU baseline \cite{garcia2008fast} and with
a state-of-the-art, sequential CPU-based algorithm.

The paper is organized as follows: in Section \ref{sec:preliminaries} we provide the formal framework used to model the processing of k-NN queries. In Section \ref{sec: knn overview} we give a brief overview about modern GPUs and review the main ideas behind \KNNGPU, as well as the main data structures used, while Section \ref{sec:pipeline knn} presents the k-NN query processing pipeline.
Finally, Sections \ref{sec:experimental setup knn} and \ref{sec:experiments knn} provide the experimental part of our work, Section \ref{sec:relworkquery processing} covers the related work while Section \ref{sec: conclusions} gives the conclusions.
Finally, Appendix \ref{app:appendix} provides further details about the complexity of the various pipeline phases.

\section{Problem Setting and Statement}
\label{sec:preliminaries}

\subsection{Problem Setting}

We consider a set of points $O=\{o_1, \ldots, o_n\}$ moving in a two-dimensional Euclidean space $\mathbb{R}^2$, where the position of
object $o_i$ is given by the function $pos_i: \mathbb{R}_{\geq 0}
\rightarrow \mathbb{R}^2$ mapping time instants into spatial positions.
These points model objects that issue position updates and 
queries as they move.  
Let ${\cal P}_i = \langle p_i^{t_0},\ldots,p_i^{t_k},\ldots \rangle$,\ $t_j < t_{j+1}$, be the
time-ordered sequence of position updates issued by $o_i$, where
$p_i^{t_j} = pos_i(t_j)$ is a position update. For a generic time $t$, $t\geq t_0$, the 
most recently known position of $o_i$ \emph{before} $t$ is denoted by $\hat{p}_i^t$, and defined as follows:
\[\hat{p}_i^t = p_i^{t_m} \in {\cal P}_i \text{ if } t_{m} < t \le t_{m+1}\]

The goal of a generic k-NN query $q_i = (x,y,k)$, is 
to find the $k$ closest objects to its center $(x,y)$.
Since we are dealing with moving objects, and the processing time $t'$ of the query 
may be later than its issuing time, we
have to consider the most recently known positions of objects.
\begin{definition}[\textsc{k-NN query}]
\label{def:knnQuery}
\textit{Let $q_i=(x,y,k)$ be a k-NN query, issued by object $o_i$, centered in $(x,y)$. Its 
 result set, computed at time $t$, is defined as:}
$$res(q_i, t) = \{ o_j \in O \ |\ rank(q_i,o_j, t) < k  
\ \land \ i \neq j   \}, $$
\textit{where} 
$$rank(q_i,o_j, t) = \left| \{o\in O\ |\ dist_t (q_i,o)<dist_t (q_i,o_j)\} \right|\ \geq\ 0$$ 
\textit{is the number of objects whose most recently known positions at time $t$ are closer to the center of $q_i$ than the position of the most recent update of $o_j$. Thus $dist_t (q_i,o_j)$ is the Euclidean distance  between the query position and the last known position $\hat{p}_j^t$ of object $o_j$ at time $t$.
}
\end{definition}

According to the above definition we observe that, when there are ties in rank comparison, the result set of a query may contain more than $k$ elements. When this happens, without loss of generality in this work we always return $k$ elements for each query by arbitrarily selecting a subset of the elements having the same maximum rank.


\hide{

Given the time-ordered sequence of queries issued by $o_i$, denoted as 
${\cal Q}_i = \langle q_i^{t_0},\ldots,q_i^{t_m},\ldots \rangle$, $t_m < t_{m+1}$ the most recent query issued by $o_i$ before time $t$, $t\geq
t_0$, is \[\hat{q}_i^t = q_i^{t_m} \in {\cal Q}_i \text{ if } t_{m} < t \le t_{m+1}\]

}

\subsection{Batch Processing}
\label{sec:batch processing}

In this work we assume that the processing of queries can be delayed up to a certain extent to optimize the overall system throughput when facing intensive workloads deriving from massive numbers of moving objects and queries, yet satisfying possible, specific QoS requirements.
We quantize the time into \emph{ticks} (time intervals) with the objective of processing updates and queries in batches 
at the end of each tick. 
Given a moving object, if we observe multiple position updates and queries issued during a time tick, only the most recent ones are processed. 
This procedure for computing queries ensures
serializable query processing and implements the timeslice query
semantics, where query results are consistent with the
database state at a given time, usually when we start processing the query \cite{gray1993transaction}.

Hereinafter we will use the following notation: $\Delta t$ is the tick duration, $\tau_k=[k \cdot \Delta t,\ \ (k+1) \cdot \Delta t)$ is the $k$-th tick, $P^{\tau_k}=\{p_1^{\tau_k}, \ldots, p_n^{\tau_k}\}$ is the set of last known positions of all objects at the beginning of the next $(k+1)$-th tick. Similarly, $Q^{\tau_k}=\{q_1^{\tau_k}, \ldots, q_n^{\tau_k}\}$ are
the most recent queries issued during $\tau_k$ and known at the beginning of the next $(k+1)$-th tick ($q_i^{\tau_k}=\bot$ in case no query was issued by $o_i$ during $\tau_k$).

\subsection{Problem Statement}
\label{sec:problem}

 
Given \emph{(i)} a set of $n$ objects $O$, \emph{(ii)} a partitioning of the time
domain into ticks $[\tau_k]_{k \in \mathbb{N}}$ of duration $\Delta
t$, \emph{(iii)} a query latency requirement $\lambda$, and \emph{(iv)} a sequence
of pairs $[(P_{\tau_k},Q_{\tau_k})]_{k \in \mathbb{N}}$, where
$P_{\tau_k}$ is the set of up-to-date object positions at the end of
$\tau_k$, and $Q_{\tau_k}$ is the set of the last issued queries
during $\tau_k$, 
the goal of \textbf{iterated batch processing} of queries $Q_{\tau_k}$ over the corresponding 
$P_{\tau_k}$, $k \in
  \mathbb{N}$, is to \textbf{compute} $[R_{\tau_k}]_{k \in \mathbb{N}}$, i.e., a \textbf{sequence of pairs}, each composed of a query and the list of the corresponding results:
\[R_{\tau_k} = \{
(q_i^{\tau_k}, \ res(q_i^{\tau_k}, \ (k+1) \cdot \Delta t)) \ \ | \ \
q_i^{\tau_k} \neq \bot \ \wedge \ q_i^{\tau_k} \in
Q_{\tau_k} \}\]

\hide
{
The \textbf{processing time} of each batch of queries $Q_{\tau_k}$ 
must be \textbf{upperbounded} as follows, to satisfy the query latency requirement
$\lambda$:
\[\mathit{processing\_time}(Q_{\tau_k}) \le
(\lambda - \Delta t) < \Delta t\]
}

\section{Proposal overview}
\label{sec: knn overview}
The use of GPUs for general purpose computation presents significant advantages, despite the more complex computational model and memory architecture. In this paper we refer to NVIDIA CUDA framework and terminology, however noting that different frameworks and architectures adopt similar solutions under different names. Here we summarize some terminology; afterwards, we discuss the main challenges related to k-NN query computation on GPUs, as well as the spatial index and the main data structures used in this work.
\subsection{Graphics Processing Units: terminology}
A GPU consists of an array of $n_{SM}$ \emph{multithreaded streaming multiprocessor} (SMs),
each with $n_{core}$ cores, yielding a total number of $n_{SM}\cdot
n_{core}$ cores. Each SM is able to run \emph{blocks} of
\emph{threads}, namely \emph{data-parallel tasks}, with threads 
in a block running concurrently on the cores of an SM.  Since a block typically has many more threads
than the cores available in a single SM, only subsets of threads, called
\emph{warps}, can run in parallel at a given time instant. Each warp consists of $sz_{warp}$ \emph{synchronous, data parallel 
threads}, executed by an SM according to a SIMD-like paradigm \cite{hong2009analytical}. We note that in current hardware a warp is managed as a block of 32 threads.

\hide{
\subsection{Graphics Processing Units: generic challenges}
Due to GPU architectural design, it is important to avoid branching inside the
same block of threads. It is worth remarking that at warp level no synchronization mechanisms are 
needed to guarantee data dependencies among threads, thanks to the underlying scheduling.
For what is related to the memory hierarchy, GPUs feature several types of memories, ranging from private 
thread registers and fast shared memory, which are both shared among the core groups
of each SM, to global memory, which has a lower throughput but it is of significant
size and represents the contact point with the CPU host. 
To achieve consistent performances a programmer has to be aware of this complex memory hierarchy by 
orchestrating and managing explicitly memory transfers between different memories, to exploit data locality whenever possible.
Finally, workload partitioning is paramount when designing GPU
algorithms, since unbalances may create inactivity bubbles
across the streaming multiprocessors which may cripple the performance.
}

\subsection{Motivating challenges and proposal sketch}

The problem of computing repeated k-NN over massive moving object observations is mainly characterized, from the GPU perspective, by the indeterminateness affecting the spatial extension of the queries, since this depends on local objects densities.
The usual approach to reduce the amount of computations per query is to adopt a spatial index based on some kind of tree. 
To compute a k-NN query one then has to perform a tree visit, exploring only parts of the tree corresponding to regions possibly enclosing nearest neighbours. 
%
%
Since the spatial extension of each k-NN query is unknown, different queries possibly require to visit different paths inside the tree or different amounts of leaves. Moreover, depending on the kind of tree used, each leaf possibly contains different amounts of objects with respect to other leaves, thus strengthening the challenge of materializing uniform GPU workloads.
In other words, the problem is to batch enough work per GPU SM while entailing uniform workloads to improve processor occupancy, limit branch divergence and ensure locality-preserving access to the GPU memory.

Our proposal tries to tackle these issues by exploiting an iterative approach in our GPU processing pipeline.
First, we adopt a spatial index based on point-region quadtrees, where blocks of close objects and queries are spatially reordered and stored according to the quadtree leaves; the leaves are in turn structurally defined by a Morton-based coding schema which permits to easily visit the tree and entails spatial locality during the processing.
Second, we iteratively process all the queries in parallel, where at each iteration the various blocks of close objects are scanned in parallel to update the result sets of corresponding blocks of k-NN queries.

\subsection{Space partitioning and indexing}
\label{sec:partitioningOverview}

\hide
{
In the context of parallel query processing, there are two main reasons for partitioning and indexing the data according to a given space partitioning approach: the first one, also common to sequential query processing, is to avoid redundant computations and access to irrelevant data, while ensuring fast access to relevant information. The second reason, 
is to exploit the index-induced data partitioning to determine balanced, independent, parallel tasks to distribute to the GPU SMs.
}

Grid-based spatial indexes are generally reported to have low construction and maintenance costs, making them suitable for update-intensive settings \cite{sowell2013experimental,sid11}, especially when exploiting GPUs \cite{lettich2014gpu}.
Among these, uniform grids may fail in guaranteeing the generation of balanced workloads when processing
skewed spatial distribution, since parallel tasks would be assigned to uneven 
blocks of objects/queries indexed by the various grid cells.
Other regular grid-based indexes, such as \emph{quadtrees}, allow achieving much better  
workload balance with relatively small construction/maintenance costs. 
Moreover, they exhibit nice mathematical/structural properties which fit quite well massively parallel architectures such as the GPUs.
Indeed, we adaptively partition the Minimum Bounding Rectangle (\textsf{MBR}) 
${\cal G}$, containing all the objects positions during any tick, 
%
into a set of variably sized cells belonging to a grid ${\cal C}$ induced by a \emph{point-region quadtree}, whose cells correspond to quadtree leaves. Each cell $c \in \mathcal{C}$ is mapped to a unique integer ID induced by a Morton coding schema, such that the total order given by IDs on cells preserves spatial locality. 
Based on the spatial partitioning ${\cal C}$, we therefore associate each object location $p \in P$ and k-NN query $q \in Q$ with the enclosing cell $c \in {\cal C}$.

\subsection{Relevant data structures}
\label{sec: knn data structures}

In this section we review the main data structures used during the query processing.

\subsubsection{General overview}
data structures containing objects or queries information use the \emph{structure of vectors} (\emph{SoV}) layout. This layout gives remarkable benefits when designing GPU algorithms, above all code reuse and efficient interplay between different operations carried on GPU \cite[Ch.33]{pharr2005gpu}. 
Moreover, it facilitates the exploitation of data locality and the use of coalescing or caching, whenever possible, thus offering substantial chances to boost the overall memory throughput, which is paramount when designing GPU algorithms.

\subsubsection{k-NN queries result set layout} 

the result set of each k-NN query (also called \emph{nearest neighbours list}) has fixed size \emph{k} and shall be stored in global memory. 
\hide
{
Since queries run and update their result lists in parallel, one might be tempted to arrange such lists by means of an interlaced layout (see for example the GPU baseline \cite{garcia2008fast}), thus favouring coalesced accesses to the GPU global memory when accessing them. However, adopting an interlaced layout with a tree-based iterative approach, as in our case, would frustrate possible benefits deriving from coalescing and caching, since the amount and the order of the updates each query has to perform is not known beforehand.
}
\begin{figure}[!h]
\centering
\includegraphics[width=0.7\columnwidth]{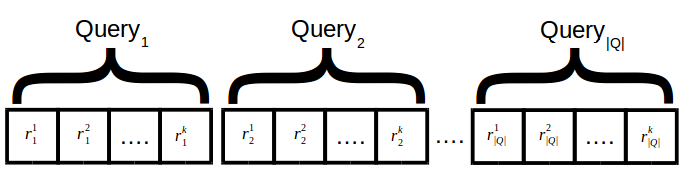}
\caption{Result set layout. $r_i^j$ denotes \emph{j}-th result of query $q_i$.}
\label{fig:linearized result set knn}
\end{figure}
We arrange such queries lists linearly, as depicted in Figure \ref{fig:linearized result set knn}. This layout, coupled with sequential accesses, proves to be quite effective in terms of exploiting caching capabilites, at least until \emph{k} does not get large. In the latter case, to still benefit from the linear layout we also propose an alternative strategy, which updates the result lists in global memory by exploiting coalesced access patterns. Indeed, when \emph{k} is large this technique turns out to be better, although it implies a slightly increased computational overhead.
We defer the details of this technique to Section \ref{sec: knn computation}.

\section{Processing Pipeline}
\label{sec:pipeline knn}


The pipeline can be described in terms of a succession of three macro stages: \emph{(i)} \emph{index creation}, \emph{(ii)} \emph{moving objects indexing}, and \emph{(iii)} \emph{iterative query processing}.
More precisely, the data entering the processing pipeline at the end of each tick are first processed to select the index parameters and build an empty index (stage \emph{(i)}). Then objects/queries  
get associated with index cells and are subsequently sorted, so that those contained within the same cell are stored in contiguous memory locations (stage \emph{(ii)}).
The subsequent phase (stage \emph{(iii)}), which essentially represents the pipeline's core, processes the k-NN queries. 

When processing repeated k-NN queries the same procedure is repeated for each tick, except 
for stage \emph{(i)} which is skipped when the objects spatial distribution does not change significantly with respect to recent, previous ticks. Thus, for the sake of readability, hereinafter we omit the subscript that indicates the tick, and denote by $P$, $Q$, and $R$, respectively, the up-to-date object positions, the 
queries, and the result set associated with a generic tick.

\subsection{Index Creation and Moving Objects Indexing}


\subsubsection{Index Creation}
we rely on a point-region (PR) quadtree as the backbone of the spatial index, exploiting PR-quadtrees ability to partition the space in differently sized parcels containing similar amounts of entities. 
Observing that - most of the times - space distributions do not change their characteristics dramatically over short time intervals, the index is rebuilt only whenever needed, i.e., when we detect that the overall amount of computations yielded during the last tick exceeds by a given factor the amount of computations yielded during past, recent ticks.
The goal of this phase, consequently, is to create - almost entirely on GPU - a space partitioning $\cal C$ over $\cal G$ where each cell of $\cal C$ is a leaf of the PR-quadtree that does not contain more than $th_{quad}$ objects. 
In the following we describe the PR-quadtree construction procedure.

First, we fix a maximum quadtree depth, $l_{max}$, and consider a \emph{virtual full quadtree} whose 
leaves are all at the same level $l_{max}$. Incidentally we observe that, in terms of space partitioning, this is equivalent to a \emph{uniform grid} of $2^{l_{max}} \times 2^{l_{max}}$ cells.
Then, we compute in parallel the Morton codes of the objects at level $l_{max}$ - this corresponds to computing the indices of the quadtree quadrants at level $l_{max}$ where objects fall - (operation carried on \textbf{GPU}) and sort the objects accordingly (by means of Radix Sort, \textbf{GPU}).
We perform such operations to exploit the direct relationship between Morton codes and quadtrees structural properties: given any Morton code $z$ associated with a quadrant $c$ at level $l_{max}$ – where c is thus identified by the pair $(l_{max}, z)$, we can determine the Morton code $z'$ of a quadrant $c'$ at any level $l \leq l_{max}$ - where $c'$ is thus
identified by the pair $(l, z')$ and $c'$ spatially includes $c$, through a simple (bitwise) operation $z'=\lfloor\frac{z}{4^{l_{max}-l}} \rfloor$.
Another property is that the initial ordering of objects obtained by sorting the Morton codes 
at level $l_{max}$ is invariant for any level $l \leq l_{max}$. In other words, the initial sorting suffices to guarantee that all the objects falling in a given quadrant of the tree (at any level $l \leq l_{max}$) are stored in contiguous memory locations.
Therefore, after the initial sorting, without loss of generality we can identify the set of object positions $P'$, $P' \subseteq P$, included in a given quadrant of the quadtree by using the \emph{interval of memory indexes} $I_{P'}=[\delta, \delta+|P'|)$ where they are actually stored.

Aftwerwards, the algorithm starts iteratively to build the \emph{actual} quadtree, level by level.
Let us suppose that $I_A$ represents the set of intervals related to quadrants which need to be split. Initially, $I_A = \{\; [0,|P|)\; \}$. Then, at each iteration, for each $(l-1)$-level quadrant added to $I_A$ for splitting, we first identify the starting/ending positions of the four intervals of the $l$-level quadtree quadrants, and store these intervals in a temporary variable (operation carried on \textbf{GPU}). Then, we determine which quadrants need further splitting at next level (their intervals are added to $I_A$, since the corresponding quadrants contain more than $th_{quad}$ objects) and which ones represent final leaves (their identifiers are added to $\cal C$) (operation conveniently carried on \textbf{CPU}). 
The process ends whenever no more quadrants need to be split, and thus $I_A$ turns out to be empty, or the maximum possible quadtree level $l_{max}$ is reached (in this case the remaining quadrants are added to $\cal C$).

The complexity of the index creation is approximatively linear in the amount of objects. We defer to the Appendix further details on the matter.

\noindent \textbf{Building a lookup table to map coordinates to cells.}
The usual approach for finding a quadtree leaf would consist in traversing the tree from the root, recursively choosing relevant nodes until the target leaf is reached. Unfortunately, on GPU this approach entails repeated irregular memory accesses and a non predictable number of operations for each leaf search. For this reason we use a different approach, characterized by a slightly larger memory footprint. 

Let us suppose that the deepest level created in a quadtree ${\cal C}$ is $l_{deep}$, $l_{deep} \leq l_{max}$. We virtually divide the space covered by ${\cal C}$ according to a uniform squared grid composed of $2^{\cdot l_{deep}} \times 2^{\cdot l_{deep}}$ cells, and denote it by ${\cal C}^{l_{deep}}$. We note that the space partitioning induced by ${\cal C}^{l_{deep}}$ also corresponds to the full quadtree having depth $l_{deep}$.
Thanks to PR-quadtree structural properties, any quadtree leaf at a level $l$, $l \leq l_{deep}$, corresponds to the union of $4^{(l_{deep} - l)}$ contiguous cells of ${\cal C}^{l_{deep}}$. 
Therefore, a mapping between ${\cal C}^{l_{deep}}$ and ${\cal C}$ cells can be easily established by means of a lookup table that realizes the function 
$z_{map}:\ {\cal C}^{l_{deep}} \rightarrow \mathcal{C}$, which associates each 
cell in ${\cal C}^{l_{deep}}$ 
with the cell in ${\cal C}$ containing it. 
This allows to retrieve the quadtree leaf that
contains any entity with simple numeric operations in constant time, making it a lightweight operation thanks to the structural simplicity underlying uniform grids.

\begin{figure}[h!]
\centering
\includegraphics[width=0.39\columnwidth]{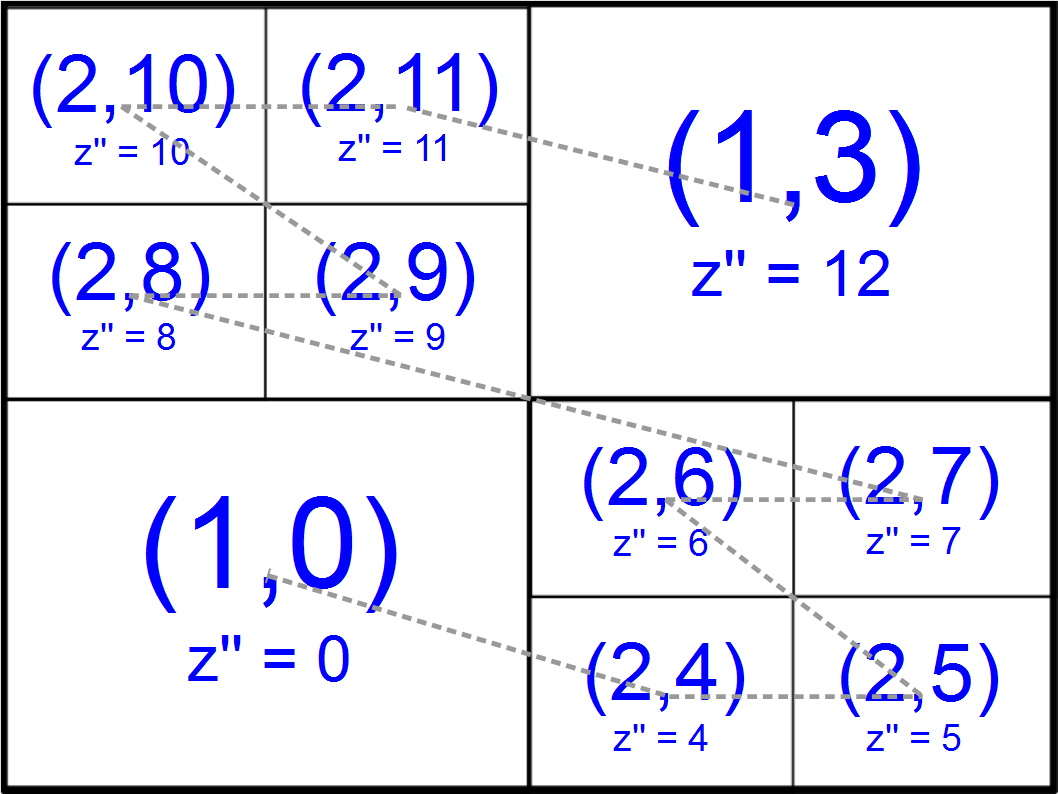}
\includegraphics[width=0.39\columnwidth]{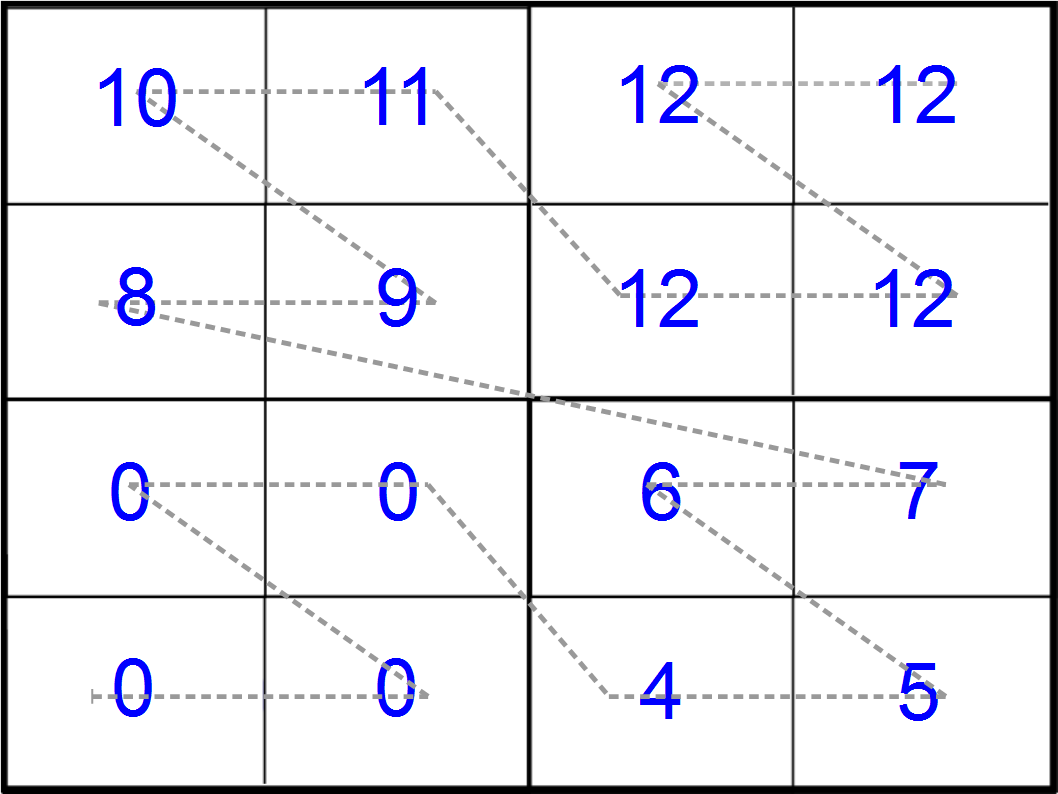}
	
\caption{Example of the mapping established by $z_{map}$ between the quadtree-induced grid $\cal C$ (left side) and the uniform grid ${\cal C}^{l_{deep}}$ (right side) related to the quadtree deepest level.}
\label{fig:zmap}
\end{figure}

At this point we can also highlight how the enumeration given to quadtree leaves actually establish a total order. This can be demonstrated by associating each leaf $(l,z)$ with the Morton code of the \emph{first} (decreasing with respect to the Morton order) quadrant it covers at level $l_{deep}$, i.e., by computing $z'' = z \cdot 4^{l - l_{deep}}$. Figure \ref{fig:zmap} shows a simple example ($l_{deep} = 2$), where the total order induced by $z''$ over the quadrants of $\cal C$ is illustrated in the left picture by the dashed line. In the right picture we see the same dashed line, yet going on at $l_{deep}$, reflecting the total order through which the identifiers of the 
${\cal C}^{l_{deep}}$'s cells are arranged in $z_{map}$.
%

%
We have to mention that the performance related to lookups in $z_{map}$ heavily depends on the ability to exploit the GPUs caching capabilities. Indeed, $z_{map}$ may have a relevant size - depending on $l_{deep}$. In light of this, it is important to design the \emph{memory layout} of $z_{map}$ to enhance data locality - this is defined by the Morton order at $l_{deep}$ - as well as to sort objects and queries according to the same order structurally defining $z_{map}$ before the actual lookups happen (see Section \ref{sec: knn objects indexing}).
%
%
%
The cost of initializing $z_{map}$ (done on \textbf{GPU}) after the index creation is, in practical terms, negligible.

\subsubsection{Moving objects indexing and active cells determination}
\label{sec: knn objects indexing}
during this phase we assign each moving object to a specific quadtree leaf. More precisely, first each object in $P$ is associated with a cell in ${\cal C}^{l_{deep}}$. Subsequently, objects are \emph{sorted} according to the identifiers of the cells to exploit caching when subsequently accessing $z_{map}$. Finally, $z_{map}$ is used  to retrieve the  identifier $(l,z)$ of the corresponding cell in the quadtree-induced ${\cal C}$. 
Finally, the resulting, sorted, struct of vectors holding $P$ is \emph{indexed} to determine the positions of the first and last object belonging to cells of ${\cal C}$ that enclose at least one object. We define 
these cells as \emph{active}, and denote the set of active cells by $\hat{\cal C}$, where $\hat{\cal C} \subseteq {\cal C}$.


The complexity of this subphase is linear with respect to the amount of objects. We defer further details to the Appendix.

\subsection{Iterative k-NN queries computation}
\label{sec: knn computation}

Once the set of active cells $\hat{\cal C}$ is determined, the actual query computation may start. Since we rely on a spatial index based on a PR-quadtree, for each query we have to perform a \emph{tree visit} to compute the list of \emph{k} nearest neighbours.
To exploit the computational power of GPUs we need to create GPU workloads such that computation is distributed uniformly across GPU SMs and objects, queries and results data are arranged so that \emph{coalescing} and \emph{caching} are exploited as much as possible. Further, since the tasks to perform are dependent on data, we have to create those workload on-the-fly, as tree visits progress. 

The underlying idea is to achieve these goals by means of an \emph{iterative} approach, exploiting spatial proximity between nearby queries as visits progress.
We distinguish between the operations carried on during the \emph{first iteration} and those carried on during \emph{subsequent iterations}.

\subsubsection{First iteration}

the first iteration orchestrates the work associated with the beginning of the queries tree visits. This corresponds to associating each query with the cell (quadtree leaf) in which its center falls, and subsequently computing the distances between the query center and the objects enclosed by the cell. The first iteration ends by updating the queries nearest neighbours lists according to the distances computed.
We structure the first iteration in two smaller phases: (i) \emph{query indexing and GPU task materialization} and (ii) \emph{distance computations} between queries and objects.

\paragraph{Query indexing and task materialization.} 
\label{par: knn query indexing task materialization}

Since each k-NN query is represented by a point, queries are mapped to specific grid cells. Consequently, such phase is equivalent to the moving objects indexing phase (Section \ref{sec: knn objects indexing}).
Since it is meaningless to consider devoid cells for the purposes of distance computations, we consider only the set of non-devoid cells containing at least a query. We denote by $\overline{\cal C}$ such set, noting that $\overline{\cal C} \subseteq \hat{\cal C} \subseteq {\cal C}$.
%
Each cell in $\overline{\cal C}$ will thus be assigned to a multithreaded task, to be scheduled to a distinct GPU SMs, where each concurrent thread is in charge of a distinct query.

Prior to the \emph{distance computations} phase, we finally  sort in descending order the set of tasks on the basis of their \emph{computational weight}, i.e., the amount of distances to be computed: by scheduling tasks according to this order we reduce possible workload imbalance across the GPU SMs.

\paragraph{Distance computations.}
\label{par: knn distance computations}

At this point the goal is to compute, for each query, the list of (up to) \emph{k} nearest objects within the assigned cell and update the lists accordingly. We saw above that the end product of the query indexing and sorting operations consists in a set of tasks, one per active cell enclosing at least one object, representing the GPU workload in charge of computing such lists.
This allows to conveniently pack together computations related to spatially nearby entities, allowing to reduce the overall amount of computations and to exploit data locality.

Accordingly, our approach relies on two pillars. The first one is represented by a GPU-friendly \emph{k-selection} algorithm based on buckets \cite{alabi2012fast}. Starting from a set of objects, this algorithm allows to find the \emph{k} nearest objects without having to explicitly store and sort distances in memory, thus reducing the overall complexity in terms of time and space.
The second pillar is represented by a proper access pattern which allows to maximize the memory throughput when updating the queries nearest neighbours lists. Considering the linear layout used for the queries result set (Figure \ref{fig:linearized result set knn}), different approaches may represent the best choice depending on \emph{k}. 

\noindent \textbf{Distance computations -- ``\emph{cached writes}" approach}. 
In this paragraph we focus on a strategy which relies only on GPU caching capabilities when updating the queries nearest neighbours lists. Algorithm \ref{lst:knn first iteration - distance computation} reports such strategy, where \textbf{local} keyword denotes automatic variables stored in private registers.

\begin{algorithm2e}[h]
\DontPrintSemicolon
\SetInd{0.7em}{0.7em}
\SetKwInOut{Input}{Input}\SetKwInOut{Output}{Output}

\Input{
\begin{itemize}[noitemsep]
\item[-] The set of active cells with at least one query, $\overline{\cal C}$.
\item[-] The reordered query set \emph{Q} and object set \emph{P}, along with the indexing information associated after  the respective sorting phases.
\item[-] The size of the queries neighbours lists, \emph{k}.
\end{itemize}
}
\Output{
\begin{itemize}[noitemsep]
\item[-] The struct of vectors containing the query result set, $(ID,DIST)$.
\item[-] The vector containing the maximum distance detected for each query, \emph{MAXDIST}.
\item[-] The vector containing the amount of nearest neighbours found for each query, \emph{NUMRES}.
\end{itemize}
}

\Begin
{
\ForEach{$c \in \overline{\cal C} \ \lstparallel{block}$}
{\nllabel{cycleBlock cache}
	\ForEach{$q \in c \ \lstparallel{thread}$}
	{\nllabel{cycleQuery cache}
		
		\BlankLine
		
		\textbf{local} $(dist_{min},dist_{max}) \leftarrow findMinMaxDist(q,c)$ \nllabel{findMinMaxDist} \;
		\textbf{local} $dist_k \leftarrow findKDist(q,c,dist_{min},dist_{max},k)$ \nllabel{findKDist} \;
		\textbf{local} $i \leftarrow 0$, $maxdist \leftarrow 0$\;
		
		\BlankLine
		
		\ForEach{$p \in c$}
		{\nllabel{cycleCopyObjects}
			\If{$((dist(p,q) < dist_k) \land (q_{id} \neq p_{id}))$}
			{
				$maxdist \leftarrow max(maxdist,dist(p,q))$\;
				$(ID_q[i],DIST_q[i]) \leftarrow (p_{id},dist(p,q))$\;
				$i \leftarrow i+1$
			}
		}\nllabel{cycleCopyObjects end}
		
		\BlankLine
		
		$MAXDIST_q = maxdist$\; \nllabel{finalWrites1}
		$NUMRES_q = i$ \nllabel{finalWrites2}
		
	} \nllabel{endCycleQuery cache}
}
}

\caption{$distComp(\overline{\cal C},Q,P,k)$}
\label{lst:knn first iteration - distance computation}
\end{algorithm2e}

Each parallel task in charge of a given $c \in \overline{\cal C}$ is assigned to a specific SM (line \ref{cycleBlock cache}) and executed according to a per-query parallelization (line \ref{cycleQuery cache}). Each thread first loads query information (line \ref{cycleQuery cache}), i.e., query coordinates and the associated cell identifier. Since queries belonging to the same cell are stored in contiguous memory locations, accesses to query data are \emph{coalesced} during the first iteration.
Subsequently, each thread finds out the minimum and maximum distance ($dist_{min}$ and $dist_{max}$ respectively) between the query center and the objects within the cell (function \emph{findMinMaxDist}, line \ref{findMinMaxDist}).  This is achieved through a simple scan over the set of objects enclosed by \emph{c}. Since every thread in a warp perform such scan by accessing objects data in the same order, and considering that different warps in a block access nearby objects, this behaviour exhibits strong locality of reference, and is thus 
suitable to fully exploit the GPU \emph{caching} capabilities.
Once $dist_{min}$ and $dist_{max}$ are determined, the algorithm goes on by finding a distance which encompasses only the \emph{k} nearest neighbours within \emph{c}, $dist_k$. This is achieved by calling the \emph{findKDist} function (line \ref{findKDist}), which implements a k-selection algorithm based on buckets \cite{alabi2012fast}, iteratively going on until a suitable distance is found.
\highlight{We note that whenever the amount of objects in \emph{c} is less than \emph{k}, the function can immediately return $dist_{k} = + \infty$ without performing any computation.}
Even inside \emph{findKDist}, threads of the same warp access \emph{c}'s objects in the same order; since different warps access objects arranged in nearby memory locations, this entails again an efficient usage of GPU caches.

Once $dist_{k}$ is determined, each thread can actually start writing sequentially the list of nearest neighbours associated with the query, according to the layout shown in Figure \ref{fig:linearized result set knn} (line \ref{cycleCopyObjects}). 
The thread terminates once it writes out the distance of the farthest object in the list, $MAXDIST_q$, and the actual amount of results written in the list, $NUMRES_q$ (lines \ref{finalWrites1} and \ref{finalWrites2}), both stored in global memory. We note that such writes are coalesced, thanks to per-query parallelization.

\noindent \textbf{Distance computations -- ``\emph{coalesced writes}" approach.}
This approach is equivalent to the one presented above except for the access pattern used to update queries lists.
More precisely, once we have computed $dist_k$ for each query, the key idea is to parallelize at warp level by assigning each query to a \emph{warp}. In turn, inside each warp we parallelize distance checks at thread level, while writes related to lists updates are cooperatively orchestrated at warp level, thus allowing to exploit coalescing when flushing out data in global memory. It is evident that such cooperative strategy requires, at \emph{task level}, the usage of a temporary buffer stored in shared memory to accumulate results, coupled with proper management operations. To this end, we exploit the native CUDA \emph{ballot} function which allows to find out (i) who found a nearest neighbour across a warp (if any), and (ii) manage writes inside the shared memory buffer without having to recur to synchronizing barriers. This approach trades a slightly increased computational complexity for a substantial increase in memory throughput when \emph{k} gets large.

\noindent \textbf{Distance computations -- Complexity.} 
The complexity of the distance computation phase is mainly dictated by the k-selection algorithm, 
whose number of iterations depends on the spatial distribution inside a cell. 
We defer to the Appendix further details.


\subsubsection{Subsequent iterations}
\label{sec: knn subsequent iterations}

after the first iteration each query is associated with a list of nearest neighbours, containing up to \emph{k} objects contained within the same cell where the query center falls. 
Depending on the objects spatial distribution and the derived index, however, it is very probable that a substantial fraction of queries have an incorrect or incomplete list. This happens when the result of a query contains at least one object falling outside the cell of the query, and such objects are nearer than the farthest object in the list computed so far. Another trivial case is when a query falls inside a cell containing less than \emph{k} objects (and $k < |P|$). Clearly, we need to look at other neighbouring cells to complete the list.
From now on, we call such queries \emph{active queries}. 
\hide
{
To better understand this issue, we present a toy example in Figure \ref{fig:knn active query}.
\begin{figure}[!h]
\centering
	\includegraphics[width=.47\columnwidth]{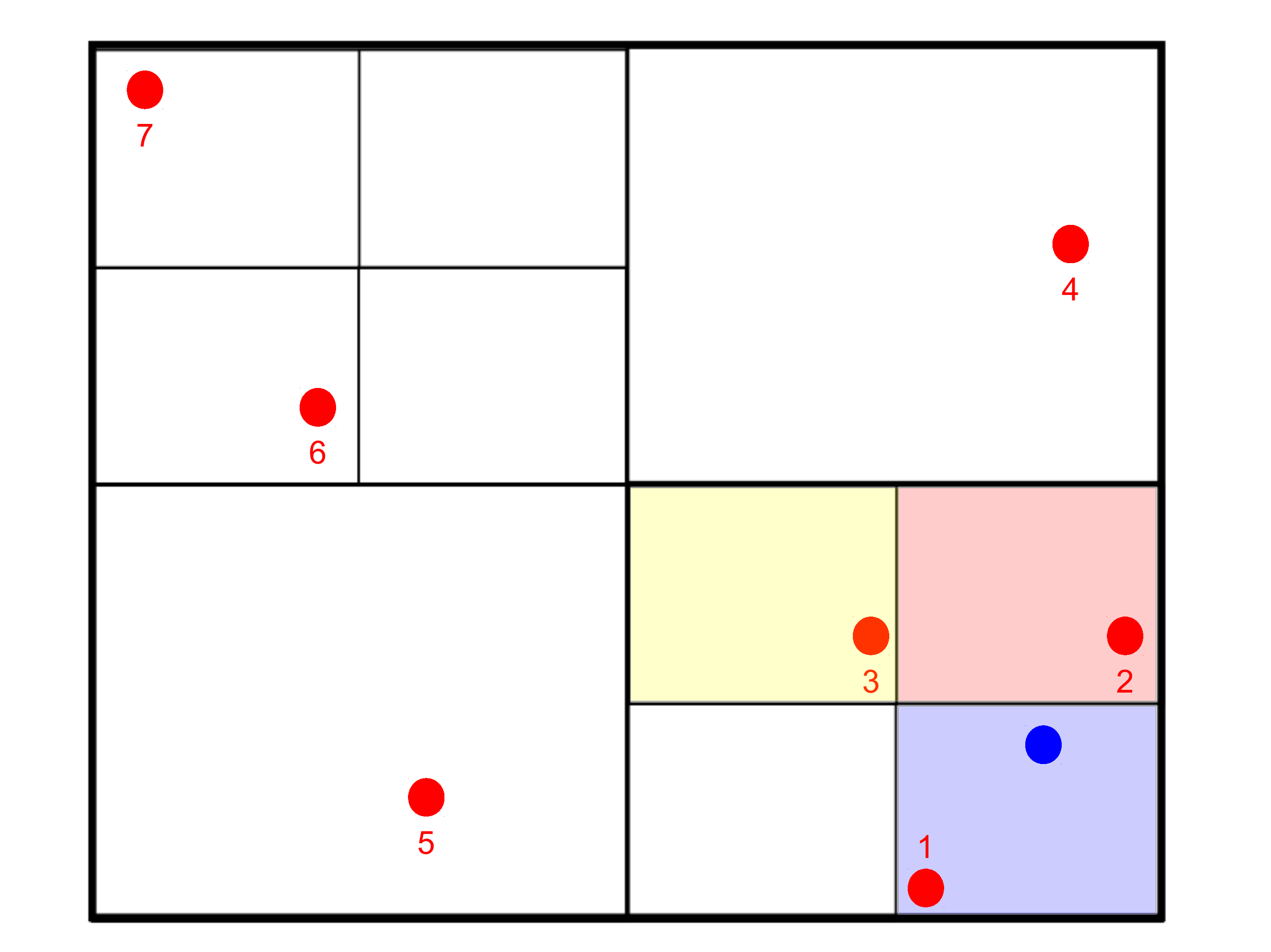}
	\hspace*{.02\columnwidth}
	\includegraphics[width=.47\columnwidth]{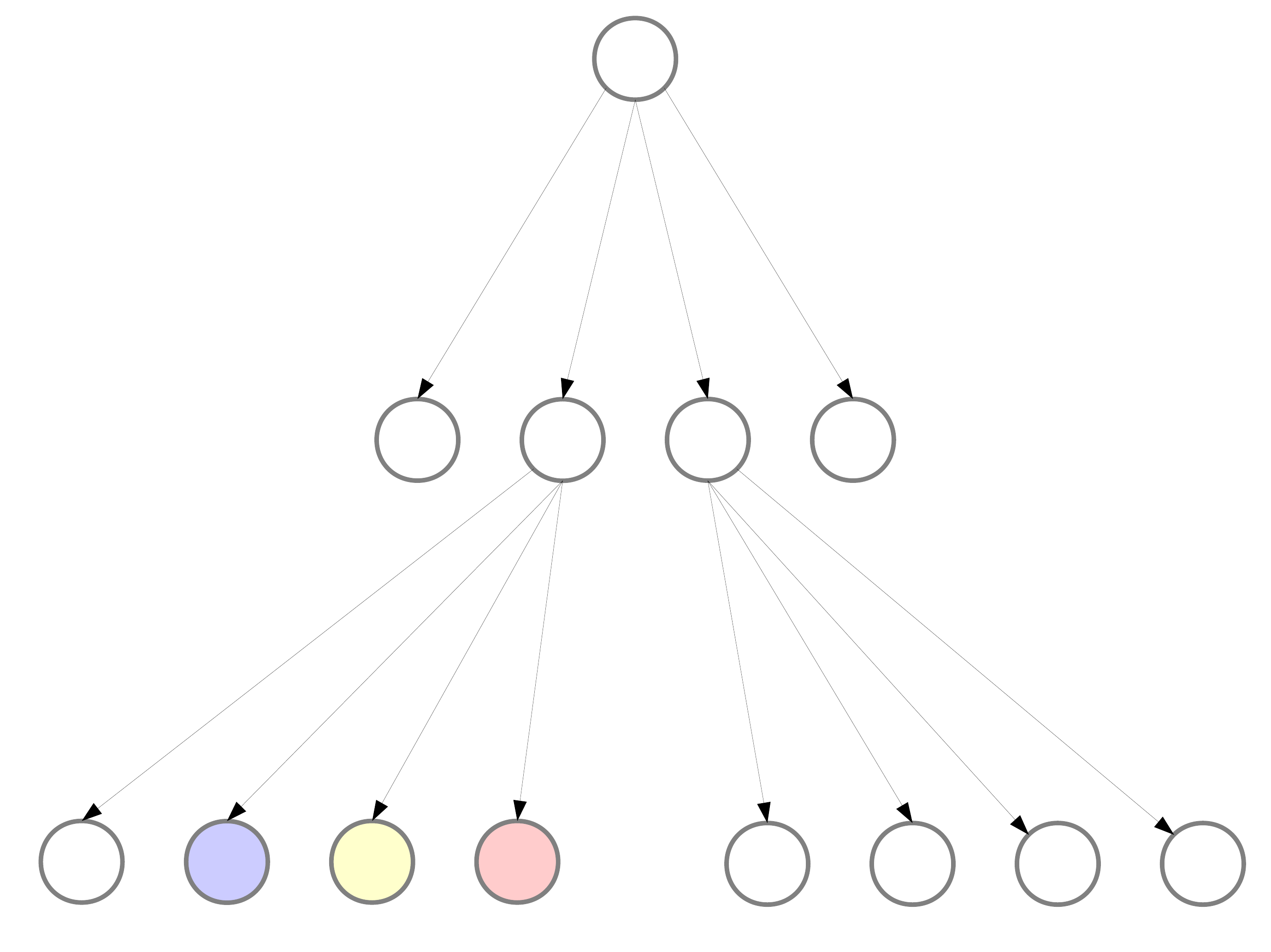}
\caption{Toy example of a 1-NN query for which we have to analyze the content of neighbouring quadrants to compute the final, correct result set.}
\label{fig:knn active query}
\end{figure}
Here we have a 1-NN query, represented by the blue dot, and seven moving objects, represented as red dots. After the first iteration we have that the list associated with the query will contain the object $\{o_1\}$, since $o_1$ falls inside the query cell (shaded in blue). However, as it is evident from the Figure objects $o_2$ and $o_3$ are nearer than $o_1$, therefore we have to consider the cells (respectively shaded in red and yellow) in which these objects fall so to compute the final, correct query list.
}
For each one of these we need to perform a tree visit, starting from the leaf in which their center fall, considering those leaves whose spatial extent may contain potential nearest neighbours, and update the query list accordingly as the visit progresses.
%
%
Considering the potential amount of active queries to be processed after the first iteration, the main challenge is to devise a strategy able to batch the work resulting from such tree visits to generate workloads suitable for GPU processing.

Our proposal is based on an iterative approach, where tree visits and distance computations related to spatially nearby active queries are packed together, iteration after iteration, until no query remains active, i.e., every query is associated with the final correct list of nearest neighbours.
At query level, the basic idea is to perform the aforementioned tree visit, starting from the leaf $c \in {\cal C}$ in which the query center falls, by splitting it into two sub-visits. Since quadtree leaves are totally ordered due to the Morton-induced encoding, as illustrated in Figure \ref{fig:zmap} (left picture), one sub-visit proceeds towards ``left'' (with respect to the order) while the other one proceeds towards ``right'', until a quadtree leaf that might contain potential nearest neighbours can be found, i.e., there is at least one non-empty leaf whose borders are \emph{nearer} than the \emph{farthest} object in the query list computed so far. The idea is to exploit the spatial locality-preserving property of Morton codes to visit, first, those leaves which likely happen to be closer to the query center. In this way, we can minimize the amount of cells to be inspected for potential objects to be returned by a query, pruning most of them on the basis of the current $dist_k$ and the borders of the inspected cells. 
To further maximize this chance of pruning, the two subvisits are conducted in an alternate fashion, 
as long as a query remains active in both directions.

In the following we sketch out our strategy in more detail.

\textbf{1 -- Iterative processing, massive tree navigation.} At the beginning of each iteration, for each query we alternatively consider one navigation direction, either left or right, to visit the quadtree leaves as long as possible nearest neighbours may be found in at least one direction, and update the navigation status of each active query accordingly.
At the end of this operation, each active query is assigned to the first unvisited leaf, in the direction considered, whose spatial extent may contain possible nearest neighbours. If no useful leaf is found along a direction, the query is flagged as \emph{inactive} with respect to that direction.


As regards the GPU implementation, the main challenge is to conveniently perform such operation by assigning each active query to a single GPU thread. Indeed, if we end up processing spatially faraway queries in the same SM, we may encounter two serious performance issues: the first one is due to \emph{execution branching} inside warps (substantial different quadtree visits may entail different flows of operation execution), while the second originates from memory issues (the quadtree information accessed by the SM threads may be stored in faraway memory locations, thus denying possible caching benefits).
To this end, we exploit the fact that active queries come already sorted from previous iterations (see Section \ref{par: knn query indexing task materialization} and the query sorting and task materialization operation described below), i.e., we assign active queries to threads according to the order achieved by means of sorting during the previous iteration and characterized by the same visit direction. Consequently, the aforementioned issues are implicitly addressed, since spatially nearby queries likely happen to be arranged in nearby memory locations due to Morton codes properties.

\hide
{
Some side remarks must be done about important design choices related to the GPU algorithm implementing this subphase: 
\textcolor{red}{first, since we check the spatial extent of quadtree quadrants when looking for possible nearest neighbours, this information can be computed by each thread on the fly without requiring to perform any lookup in memory. 
Second, we always reach $l_{deep}$ before possibly assigning a query to a new quadtree leaf through the inverted index $z_{map}$: even if this may slightly elongate tree traversals, this avoids to perform a lookup in memory each time we need to check whether any quadrant at any level is a leaf or not.}
}

At thread level, we have to update the navigation status of a query restarting from the \emph{previously assigned} leaf. To this end, we consider again the virtual full-quadtree having depth $l_{deep}$, ${\cal C}_{l_{deep}}$, instead of the actual one. In this sense, we (possibly) assign a quadtree $\cal C$ leaf to a query only whenever we detect a quadrant at $l_{deep}$ whose borders are closer than the farthest object in the query list (at distance $dist_k$ from the query). If this happens, we then just need to perform a single lookup in $z_{map}$, similarly to what is done in the moving objects and query indexing phases, to retrieve the enclosing  leaf of $\cal C$.
This design choice gives a big major advantage from the GPU perspective, in that the virtual full-quadtree, thanks to its structural regularity, can be navigated without performing any lookup in memory, since the spatial extension of its quadrants can be determined mathematically on-the-fly by each thread. The only lookup needed is the one we have to perform when we detect a quadrant of interest in $l_{deep}$. Secondarily, since queries are assigned to threads according to the sorting order yielded during the previous iteration characterized by the same visit direction, we exploit data locality when accessing $z_{map}$.

Query-wise, the complexity of this subphase is dictated by the amount of active queries in the considered direction. At query level, the worst case scenario corresponds to visiting all ${\cal C}_{l_{deep}}$ nodes.

\textbf{2 -- Iterative processing, query sorting and task materialization.} Subsequently, queries are \emph{sorted} according to the newly assigned cell identifiers. The same sorting operation also partitions between active and inactive queries as well: once active queries get sorted it suffices to find the first query having the inactive flag set to determine the extent of both sets. Such simple, yet massively parallel, operation is conveniently performed on GPU.

After the sorting operation we only focus on those queries still considered active. By virtue of sorting, the
queries 
associated with the same cell are displaced in contiguous memory locations. We exploit again this property, as in Section \ref{par: knn query indexing task materialization}, to determine the first and last query for each cell having at least one query assigned. The outcome is represented again by a set of tasks, one per active cell associated with at least one active query.  We denote such set by $\overline{\cal C}$, as done previously.

\textbf{3 -- Iterative processing, nearest neighbours lists update.} Afterwards, we proceed by updating the result lists of all the active queries by considering their newly assigned index cells. 
This step is almost equivalent to the one described in Algorithm \ref{lst:knn first iteration - distance computation}: for each query, first we find out the new $dist_k$, by considering the objects contained in the current result list of the query and those occurring in the newly assigned cell (this is done again by means of the k-selection algorithm).
Finally, we update the list according to the new $dist_k$. Here it is worth remarking that for each query we can \emph{prune} out consistent amounts of objects in the newly assigned cell simply by looking at the distance of the farthest object in the result list found so far. Obviously, such optimization cannot be used in case a query has less than \emph{k} neighbours in the list.
The complexity of this subphase is similar to the \textbf{Distance computation} one, performed during the first iteration (Section \ref{par: knn distance computations}). We defer further details to the Appendix.

\section{Experimental Setup}
\label{sec:experimental setup knn}

All the experiments are conducted on a PC equipped with an Intel Core i7 2600 CPU, running at
3,4 GHz, with 16 GB RAM and a Nvidia GTX 580 GPU with 3 GB of RAM coupled with CUDA 5.5.
The OS is Ubuntu 12.04.
We exploit a publicly available framework \cite{sowell2013experimental} for both workload generation and testing. We use three types of synthetic datasets: \textit{uniform datasets}, in which moving objects are distributed uniformly in the space, \textit{gaussian datasets}, in which moving objects gather around \textit{hotspots} by following a normal distribution, and \textit{road network datasets}, where objects are distributed uniformly over the edges of a network (in our case, the San Francisco road network) and move along the edges. In the gaussian datasets case the skewness depends on the amount of hotspots: the more they are, the more the resulting distribution is uniform.
In all tests we compute repeated k-NN queries over 30 ticks; \highlight{each query is issued by a moving object - according to the problem statement introduced in Section \ref{sec:problem}.} To model object movements the framework generates 30 instances of each dataset, one for each tick. \highlight{The overall amount of queries per tick is equal to the amount of moving objects (i.e., one query per object)}. Table \ref{tab:param knn} summarizes the main parameters used to generate the datasets. The framework uses a generic spatial distance unit $u$ (e.g., meters).
\highlight{The usage of synthetic (or partly synthetic) datasets only is consistent with previous literature \cite{sowell2013experimental} and can be motivated by observing - apart from the lack of suitable real-world datasets - that the datasets characteristics space possibly influencing the algorithms performance is very vast; as a consequence, exploring effectively such space is feasible only if one has full control over such factors.}

We use \KNNGPU\ to denote our proposal. We denote by \KNNGPUCACHE\ the flavour using the \emph{cached writes} approach, while \KNNGPUCOALESCE\ uses the \emph{coalesced writes} approach. Wherever not specified, \KNNGPU\ = \KNNGPUCACHE.
As CPU sequential competitor we consider the sequential k-NN search algorithm offered by the well-established FLANN library
\footnote{http://www.cs.ubc.ca/research/flann/, version 1.8.4.} and is denoted by \KNNCPU. 
The GPU-based baseline is the brute-force algorithm presented in \cite{garcia2008fast} and is denoted by \KNNBASELINE.

\begin{table}[h!]
\centering
\small
\begin{tabular}{|p{.2\columnwidth}|p{.7\columnwidth}|}
  \hline\hline
  \emph{Spatial region} & 
  All tests occur in a squared spatial region with side length of $22500\ u$.
  \\ \hline
  \emph{Objects maximum speed} &
  In all tests the maximum speed of each object is fixed to $200\ u$ per tick ($\Delta t$), where the objects are allowed to change their speed as described in \cite{sowell2013experimental}. In general, changes in speed may slightly alter the objects distribution but do not change the distribution general properties.
  \\ \hline
  \emph{Query rate} &  The percentage of objects that issue a k-NN query during every tick is always set to 100\%.
  \\ \hline\hline
\end{tabular}
\caption{\label{tab:param knn}
Relevant data and workload generation parameters.}
\end{table}

\section{Experimental Evaluation}
\label{sec:experiments knn}

The experimental studies conducted for this work are the following:

\begin{itemize}
\item \emph{S1}: We study how \KNNGPU 's performance is affected when varying the maximum amount of objects per quadtree leaf (thus influencing the tree height), the amount of results per query (\emph{k}), and the dataset skewness.
\item \emph{S2}: We compare \KNNGPU\ against \KNNBASELINE.
\item \emph{S3}: We compare \KNNGPU\ against \KNNCPU.
\end{itemize}

\subsection*{S1 -- Tree height, neighbours list size \textit{k} and spatial skewness impacts on \KNNGPU 's performance}
\label{sec: Study S1 knn}

\noindent \textbf{Tree height and neighbours list size.}
When processing any dataset, two crucial parameters are represented by the tree height, which is controlled indirectly by altering the maximum amount of objects admitted in a single quadtree leaf ($th_{quad}$), and the neighbours list size \emph{k} associated with the each query. Choosing an optimal $th_{quad}$ is strictly connected to \emph{k}: if the tree height is too high with respect to \textit{k}, then many leaves with few objects shall be visited, thus increasing the amount of operations needed to perform such operations. On the other hand, if the tree height is too low we end up visiting few leaves with many objects, thus possibly performing a relevant amount of useless computations due to the reduced pruning power of the index. As a consequence, it is necessary to find an appropriate $th_{quad}$ for a given $k$.

\begin{figure}[h!]
    \centering
    \includegraphics[width=0.49\columnwidth]{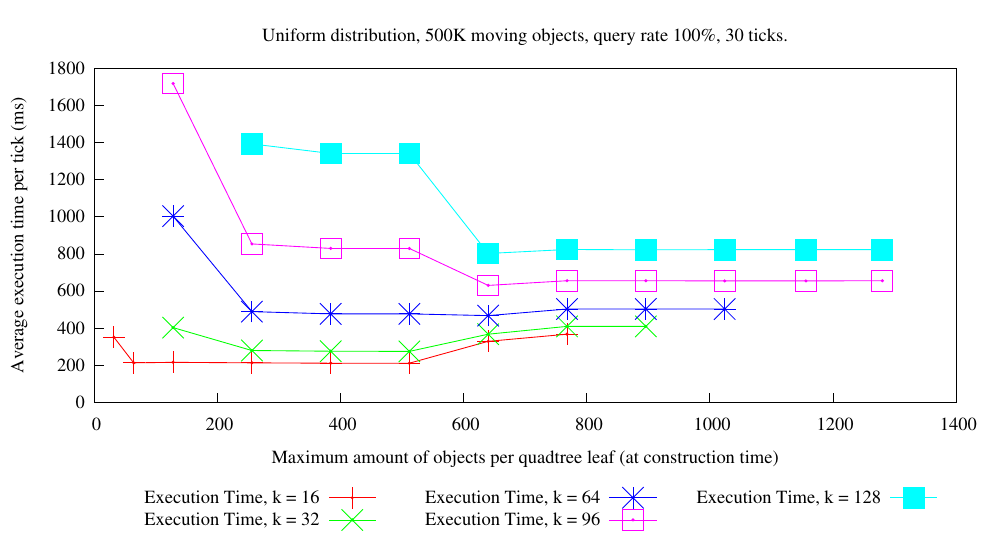}
    \includegraphics[width=0.49\columnwidth]{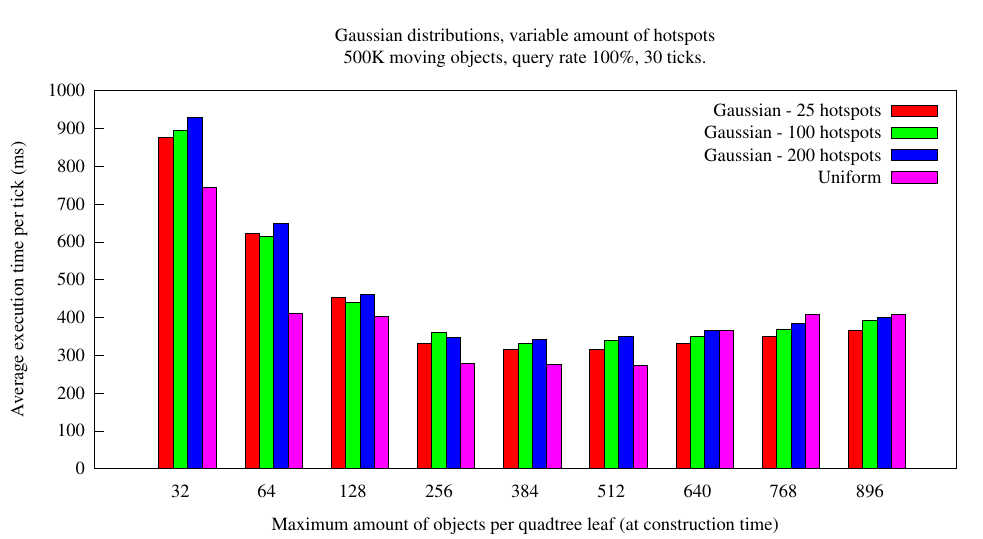}
    \caption{\textit{Left plot}: Relationship between tree height (indirectly controlled) and \emph{k}, and its repercussions on \KNNGPU 's performance. \textit{Right plot}: Skewness repercussions on execution time and $th_{quad}$'s optimality.}
    \label{fig:s1 exp knn}
\end{figure}

Figure \ref{fig:s1 exp knn}, \emph{left plot}, shows how each \emph{k} is associated with a range of optimal $th_{quad}$ values for which the algorithm's execution time is minimized. We observe how such ranges are relatively wide as well, a property which is desirable since this minimizes performance fluctuations even when not using an optimal $th_{quad}$. 
Finally, even if it is not evident for every curve the execution time starts  increasing again whenever $th_{quad}$ gets too high with respect to \emph{k}, since the pruning power of the index gets reduced. Incidentally, we observe how the execution time increases whenever \emph{k} increase. \highlight{We note that the same performance trends can be replicated with skewed distributions (such as gaussian and network ones), even though characterized by slightly higher execution times.}\\
\textbf{Spatial skewness impact on finding an optimal $th_{quad}$.}
In this batch of experiments we want to check whether the skewness has relevant impacts on finding an optimal $th_{quad}$ range. 
Datasets are distributed according to a gaussian distribution, each characterized by a different amount of hotspots to yield differently skewed distributions. All datasets have a fixed amount of objects equal to 500K; other dataset characteristics are set to their defaults, according to Table \ref{tab:param knn}.
The size of nearest neighbours lists is set to $k=32$.
From Figure \ref{fig:s1 exp knn}, \emph{right plot}, we see how the skewness does not influence, if not marginally, the optimal $th_{quad}$ range for a given \emph{k}.
Even if in this series of experiments we use a fixed \emph{k}, it is possible to show that the performance trends associated with different \emph{k} values are approximately the same.

\hide
{
\paragraph*{Query rate}

In this batch of experiments we want to determine the impact of the query rate on \KNNGPU 's performance.
In the following we use a fixed uniform dataset having 500K objects; other dataset characteristics are set according to defaults (Table \ref{tab:param knn}). We set the amount of nearest neighbours per query to $k = 32$; accordingly, we set $th_{quad} = 384$ to guarantee that \KNNGPU\ exhibits the best possible performances.

\begin{figure}[h!]
    \centering
    \includegraphics[width=\columnwidth]{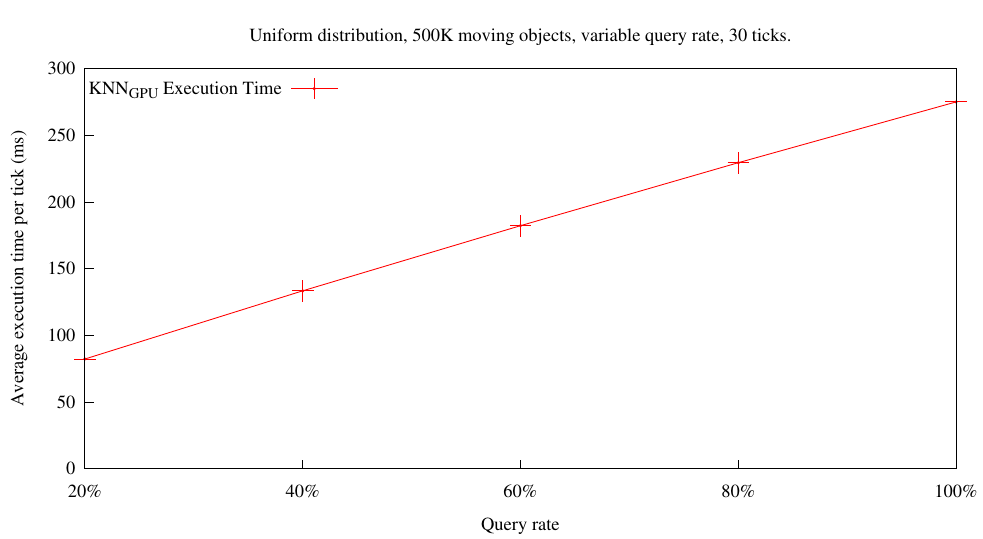}
    \caption{Query rate influence on \KNNGPU 's performance. The amount of objects (500K), the neighbours list size ($k=32$) and the maximum amount of objects per quadtree leaf ($th_{quad} = 384$) are all fixed across the experiments.}
    \label{fig:s1 exp3 knn}
\end{figure} 

From Figure \ref{fig:s1 exp3 knn} we see how the query rate influences linearly \KNNGPU 's execution time, as expected from the complexities described in Section \ref{sec:pipeline knn}.
}

\subsection*{S2 -- \KNNGPU\ vs \KNNBASELINE}
\label{sec: Study S2 knn}

In this study we take into consideration two main parameters, i.e., the \emph{amount of objects} and the amount of nearest neighbours per query, \emph{k}.
In light of the results shown in study S1, for what regards \KNNGPU\ we set $th_{quad} = 12k$ when $32 \leq k \leq 256$ since this assures, on average, the best performance, while we use $th_{quad} = 192, k < 32$ and $th_{quad} = 2048, k > 128$.

In the first batch of experiments we consider a uniform distribution, where we vary the amount of moving objects and set $k=32$, while other dataset characteristics are set according to defaults.
\begin{figure}[h!]
    \centering
    \includegraphics[width=0.49\columnwidth]{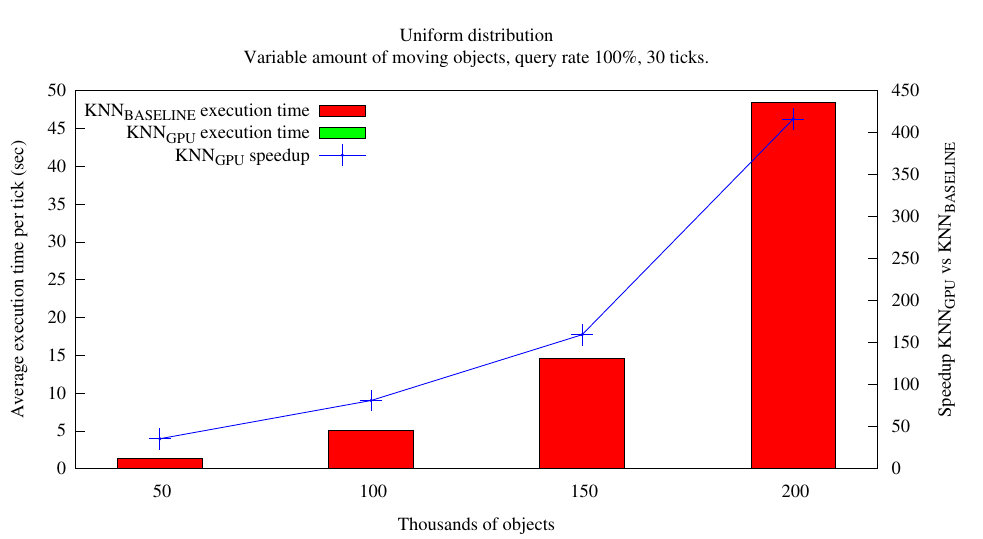}
    \includegraphics[width=0.49\columnwidth]{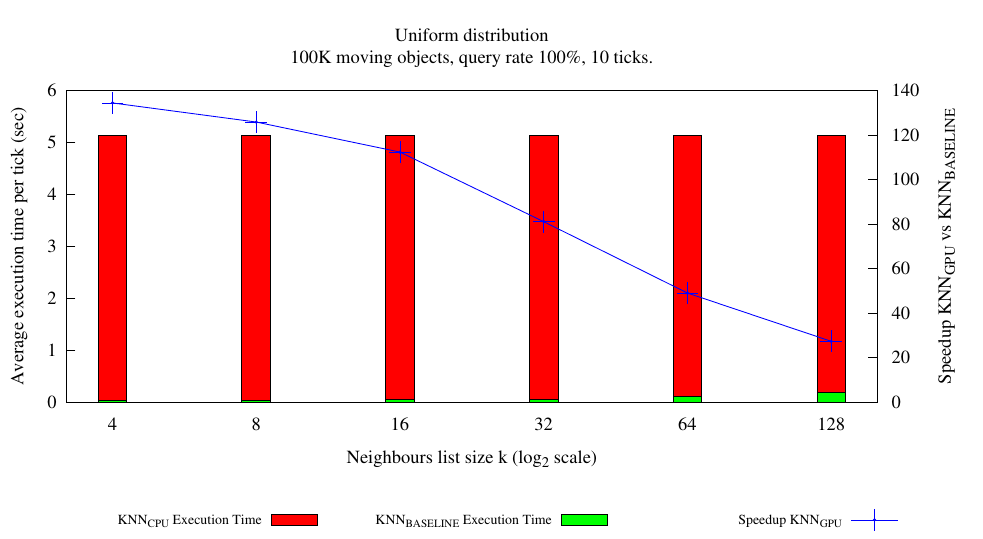}
    \caption{\KNNGPU\ vs \KNNBASELINE. \textit{Left plot}: variable amount of moving objects, $k=32$. \textit{Right plot}: variable nearest neighbours list size \emph{k}, fixed amount of objects (100K).}
    \label{fig:s2 exp knn}
\end{figure}
From Figure \ref{fig:s2 exp knn}, \emph{left plot}, we see how \KNNGPU\ heavily outperforms \KNNBASELINE\ as soon as the amount of objects gets relevant. In the second batch of experiments we still use a uniform distribution and vary the amount of nearest neighbours per query (\emph{k}) while keeping fixed the amount of moving objects (100K). Other characteristics are set to defaults.
From Figure \ref{fig:s2 exp knn}, \emph{right plot}, we see how \KNNBASELINE's execution time is fixed, since it depends exclusively on the amount of distances to compute and sort, while \KNNGPU's execution time increases when \emph{k} increases - as expected - thus reducing its advantage gap.

\subsection*{S3 -- \KNNGPU\ vs \KNNCPU}
\label{sec: Study S3 knn}

In this section we compare \KNNGPU\ against \KNNCPU. \KNNCPU\ relies on kd-trees for computing k-NN queries. In our tests we force \KNNCPU\ to use 1 CPU core and an optimized L2 distance functor. We also set to 32 the maximum amount of objects per kd-tree leaf, since this value proves to be the best choice in our experimental setting.
For what regards \KNNGPU, we use the same settings used in S2 for $th_{quad}$.
In this analysis we consider datasets characterized by different spatial distributions and different amounts of moving objects. Also, we study how \KNNGPU\ behaves according to different amounts of nearest neighbours per query (\emph{k}).

\noindent \textbf{Varying the amount of moving objects.} 
In the following batch of experiments we study  \KNNGPU\ and \KNNCPU\ when varying the amount of moving objects. The dataset characteristics varied across the experiments are (i) the spatial distribution and (ii) the amount of moving objects (between 100K and 1500K). In the Gaussian case the amount of hotspots is fixed to 25, thus yielding a moderately skewed distribution, while other characteristics are set to defaults.

\begin{figure}[h!]
    \centering
    \includegraphics[width=0.49\columnwidth]{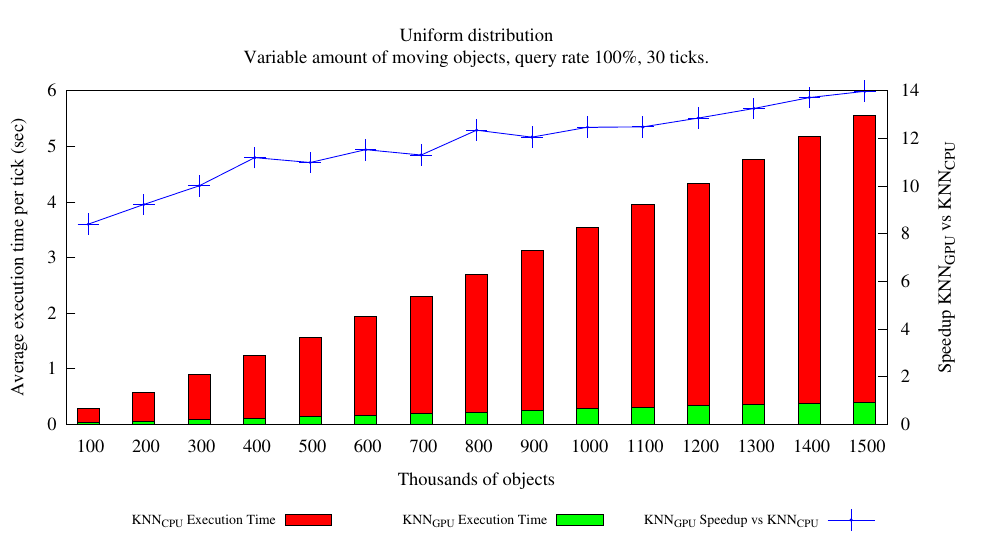}
    \includegraphics[width=0.49\columnwidth]{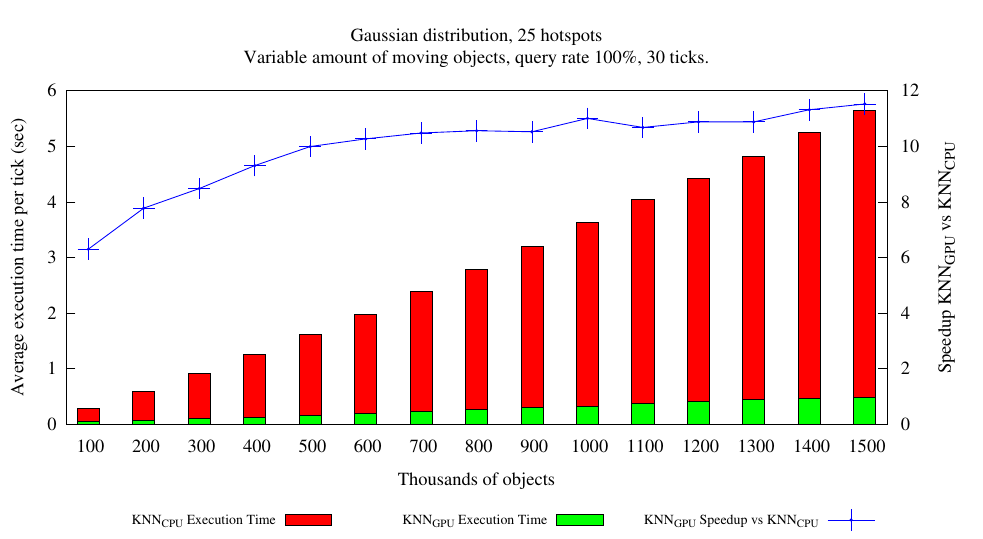}
    \includegraphics[width=0.49\columnwidth]{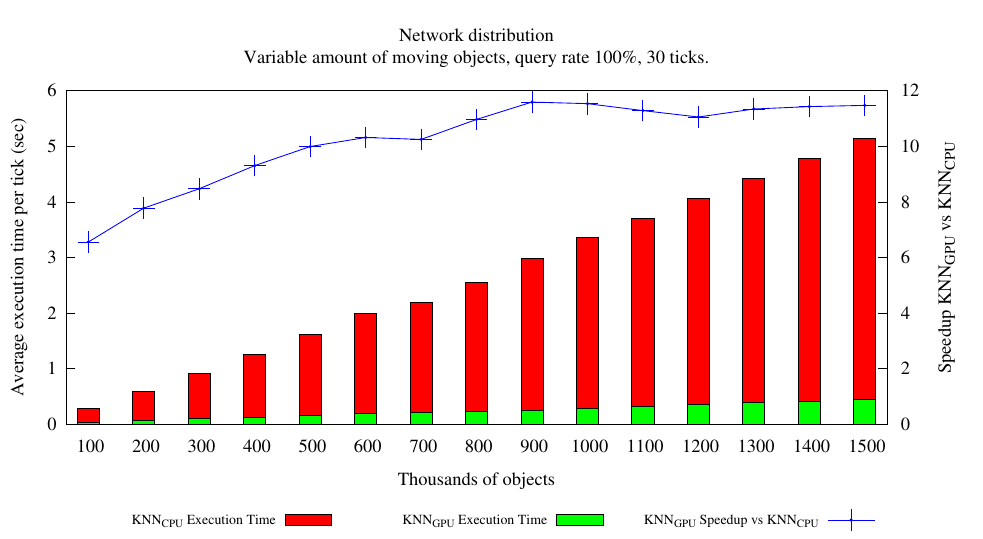}
    \caption{\KNNGPU vs \KNNCPU, variable amount of objects, fixed $k=32$.}
    \label{fig:s3 exp1 knn}
\end{figure}

From Figure \ref{fig:s3 exp1 knn} we see how, in general, \KNNGPU's speedup increases whenever the amount of objects increases, since the amount of calculations (in terms of distances to compute) increases, thus making the query processing more and more a compute-intensive task - therefore favouring \KNNGPU. 
The other main observation relates to the skewness: speedups achieved with skewed distributions are slightly lower than those achieved with uniform distributions; this fact is expected, since the indexing used by \KNNGPU\ cannot totally avoid imbalances between single tasks. 

\noindent \textbf{Varying the neighbours lists size, \emph{k}.} 
In this batch of experiments we study how \KNNGPU compares with respect to \KNNCPU\ when varying the neighbours list size \emph{k} and the spatial distribution. 
This time we use a fixed amount of objects, 1 million, while varying \emph{k} in the [1,512] range. All other dataset characteristics are set to defaults.
\begin{figure}[h!]
    \centering
    \includegraphics[width=0.49\columnwidth]{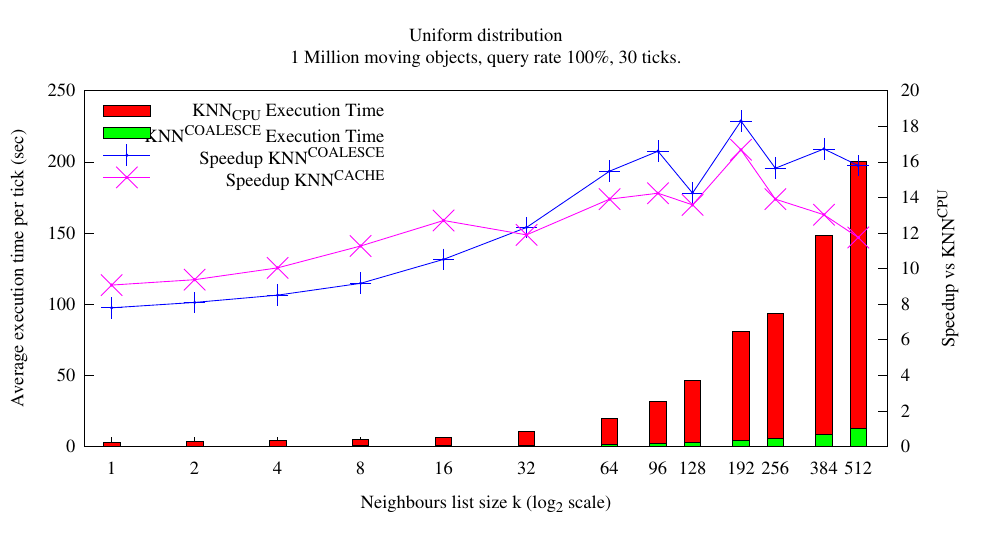}
    \includegraphics[width=0.49\columnwidth]{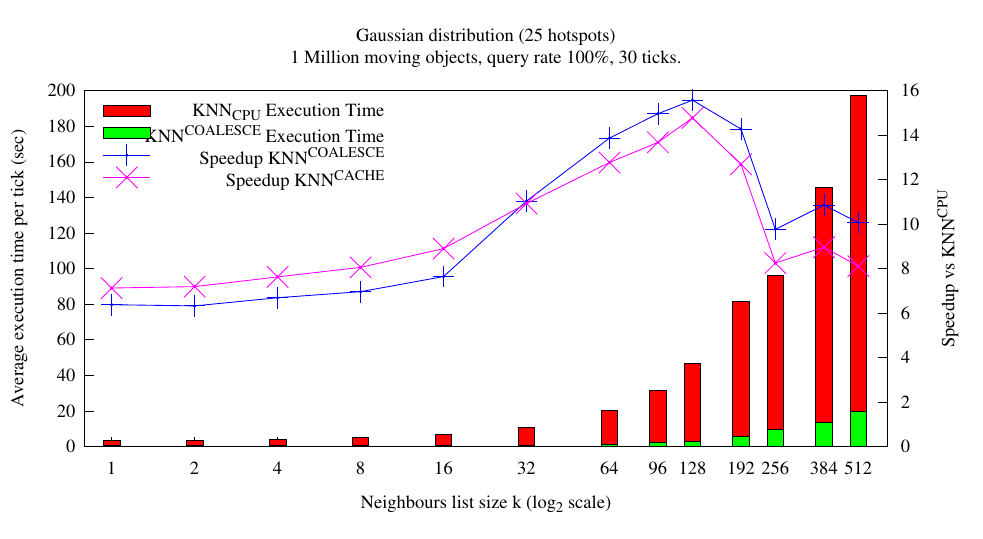}
    \includegraphics[width=0.49\columnwidth]{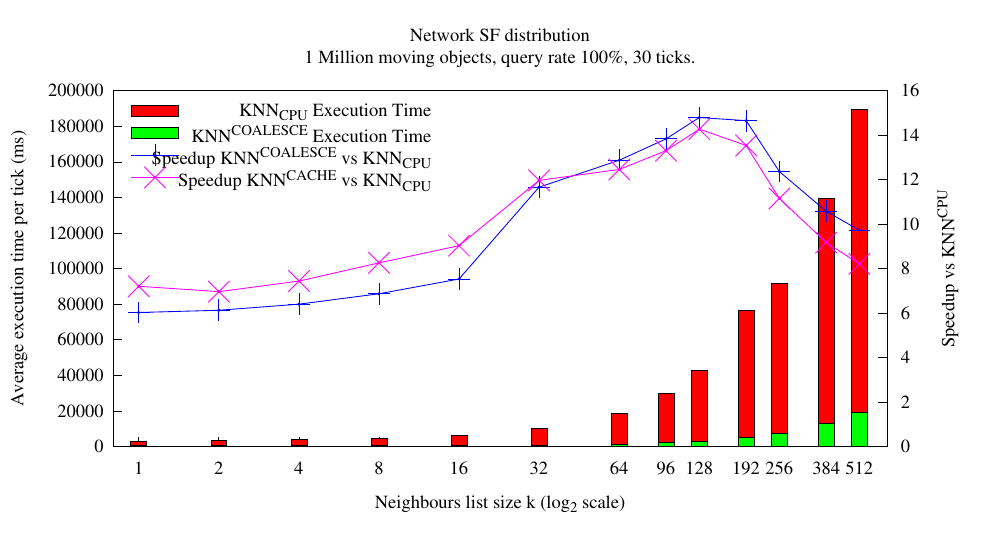}
    \caption{\KNNGPUCOALESCE\ vs \KNNGPUCACHE\ vs \KNNCPU, variable \emph{k}.}
    \label{fig:s3 exp2 knn}
\end{figure}

From Figure \ref{fig:s3 exp2 knn} (in the plot the default variant used is denoted by \KNNGPUCACHE) we see how increasing \emph{k} up to a certain point has positive effects on the performances, since the processing becomes more and more compute-intensive while GPU caching is still effective. 
However, when \emph{k} gets very high we observe performance degradation due to a decreased GPU caching efficiency, which is expected considering the linear layout (Figure \ref{fig:linearized result set knn}) used for the queries result set.
%
%
For what relates to the skewness we observe again how it mildly influences negatively \KNNGPU 's performances with respect to uniform distributions.\\
\textbf{Improving the memory throughput when \emph{k} gets high}: from Figure \ref{fig:s3 exp2 knn} we observe how \KNNGPUCOALESCE\ outperforms \KNNGPUCACHE\ whenever \emph{k} is equal or greater than the size of a warp ($k \geq 32$), while we have the opposite whenever \emph{k} is lower.
These results can be explained by observing some key facts. First, \KNNGPUCOALESCE's strategy becomes effective once threads inside a warp get fully utilized and the required amount of nearest neighbours per query gets more and more relevant, thus making \KNNGPUCACHE's strategy less effective in terms of memory throughput. Indeed, this combination of factors is reflected in the performance trend observed in all the figures.
Second, to exploit coalescing we need a proper access pattern, which in turn requires some computational overhead to orchestrate the computations accordingly. Whenever \emph{k} is low such overhead, coupled with a slight thread under-utilization when updating the queries lists, slightly penalizes \KNNGPUCOALESCE.

\section{Related work}
\label{sec:relworkquery processing}

\hide
{
The problem of computing a given set of k-NN queries over a given set of points in some \emph{d}-dimensional space is a well-renowned problem found in many theoretical and practical fields. The general schema used to tackle this problem usually follows a two step approach: first, some \emph{spatial index} is built over the set of points to reduce the amount of computations per query with a subsequent \emph{recursive search} phase, which is performed over the index to compute the query results.
In general, the effectiveness of a solution depends on the underlying spatial index and on the relaxation of some problem constraints related to the quality of results.
}
In the following we review relevant works concerning low dimensional spaces, since this work considers $\mathbb{R}^2$.
For what is related to traditional architectures, the vast majority of the approaches are based on the usage of kd-trees, as shown in extensive surveys such as \cite[Ch. 63]{mehta2004handbook} or \cite[Sec. 5]{Chavez01}, while other solutions are based on R-Trees \cite{samet99}.

To date, the first work tackling the problem of computing k-NN queries by means of a hybrid CPU/GPU approach is \cite{garcia2008fast}, where the authors propose a brute-force, quadratic, approach.
This approach is quite simple and straightforward, yet it fits quite well the GPUs architectures and proves to be quite effective with small/medium sized datasets having high dimensionality. \highlight{An improved brute-force strategy is presented in \cite{sismanis12}, where the authors use a variant of the bitonic sort algorithm, which keeps into account \emph{k} when sorting the elements, to significantly reduce the overall execution time.}
Focusing specifically on works related to low dimensional spaces, many actually tackle the problem in static scenarios and when the problem is part of a bigger problem. Indeed, computing k-NN queries is strikingly recurrent in many fields, such as computer graphics, physics, astronomy, etc. (e.g., \cite{heinermann2013gpu, nakasato2012implementation, qiu2009gpu, popov2007stackless}). 
Such works almost always rely on a kd-tree based index, whose construction usually happens on CPU.
The index is subsequently used to perform the k-NN search, where the tree has to be navigated for each query: such navigation is usually performed on CPU, depending on the specific approach, while distance computations always happen on GPU.
All these solutions typically exhibit a bottleneck, in that at least one core operation is performed on CPU whereas parallelizing on GPU would speed up the entire processing remarkably; moreover, these works are tailored for very specific scenarios.
%
\highlight{Other few works tackle problems more similar to the one presented here. 
For instance, \cite{fort2009gpu} proposes an approach which uses the GPU to speed up the computation of k-NN queries over a set of static points dislocated on a given road network. The approach uses a spatial index derived from Voronoi diagrams, which turns out to be computationally expensive to compute and lookup on GPU and is therefore impractical when tackling massive amounts of points or queries.
Another work \cite{wei2013parallel} considers a (possibly massive) set of moving objects dislocated along a road network and, equivalently to our work, it tackles the problem of computing k-NN queries issued by a fraction of the same objects. Since the computation of the distances depends on the topology of the network, the main idea is to use the network as the underlying spatial index to limit the overall amount of computations per k-NN query. The experimental part of this work seems to indicate that the proposed approach is effective only when processing very moderate amounts of queries.}

A recent work \cite{gieseke2014buffer} targets spaces having low to moderate dimensionality (i.e., $\mathbb{R}^{4 \leq d \leq 20}$). This work is quite interesting in that, starting from a spatial index based on kd-trees, it exploits queue-based buffers to accumulate and distribute on the fly fairly uniform workloads across GPU streaming multiprocessors. Analogously to our proposal, these workloads derive from tree visits bounded to k-NN queries.

\section{Conclusions}
\label{sec: conclusions}

In this paper we presented a novel hybrid CPU/GPU query processing pipeline, relying on scalable grid-based spatial indexes and on an iterative approach - coupled with ad-hoc data structures and proper memory access patterns, capable of computing massive amounts of repeated k-NN queries over massive moving objects observations. The solution proposed is the first known to effectively exploit the GPUs architectural features to speed-up the query processing in such scenarios and, at the same time, contend effectively with skewed spatial distributions of objects and queries.
We extensively tested our solution to study its sensitivity to parameters and data distribution. In these experiments we also prove that our solution outperforms a baseline GPU approach and achieves significant performance gains over a state-of-the-art CPU sequential competitor.

\bibliographystyle{unsrt}
\bibliography{knn}

\break
\appendix

\section{Additional details}
\label{app:appendix}

In the following we review in greater detail the main complexities characterizing the various phases of the processing pipeline.

\subsection{Index construction}
\label{app: index construction complexity}

In Section \ref{sec:batch processing} we introduced the index construction phase. Here we detail the complexity characterizing this operation.
The computation of single Morton codes has a fixed cost determined by the number of bits used for coordinate representation; therefore, computing Morton codes in parallel for all objects has a complexity equal to $O(|P|)$. 
Since Radix Sort \cite{merrill11} is used for the subsequent sorting operation, the complexity related to sorting is $O(b \cdot |P|)$, where $b$ represents the base value used during sorting: given that $b \ll |P|$, the overall complexity can be approximated to $O(|P|)$.

After the sorting operation, the actual iterative PR-quadtree construction begins. The operation related to detecting quadrants which need to be split has a worst-case complexity equal to 
$$O \bigg( l_{max} \cdot |P| + 2\sum_{l=0}^{l_{max}}4^l \bigg),$$
where the first term is due to the amount of objects scanned at each iteration, while $2 \cdot \sum_{l=0}^{l_{max}}4^l = 2 \cdot \frac{1-4^{l_{max}-1}}{1-4}$ represents the maximum amount of starting and ending indices related to the $4^l$ quadtree quadrants at any level \textit{l}. Since $4^l \ll |P|$, the related computational overhead is negligible and therefore the average complexity can be safely approximated to $O(l_{max} \cdot |P| + 2\sum_{l=0}^{l_{max}}4^l) \simeq O(l_{max} \cdot |P|)$. 
The remaining operation, carried on CPU, in which we determine which quadrants need to be split at the next level has a complexity equal to $2\sum_{l=0}^{l_{max}}4^l$ and, consequently, has negligible cost. 

In conclusion, since we impose $l_{max}$ to be a low constant (e.g., $l_{max} \leq 10$) and the amount of quadrants created per each level is orders of magnitude lower than the amount of objects, the overall complexity of the index creation phase can be approximated to $O(|P|)$.

\subsection{Lookup table $z_{map}$ initialization}
\label{app:zmap init complexity}

In Section \ref{sec:batch processing} we introduced the lookup table $z_{map}$ used to retrieve in constant time the identifier of the quadtree leaf in which an object/query falls. Here we detail the complexity related to its initialization, which happens immediately after the index construction phase. 

This operation is performed entirely on GPU by assigning each ${\cal C}$ cell (quadtree leaf) to a GPU streaming multiprocessor, which in turn initializes in parallel the interval of cells (elements of the lookup table) in ${\cal C}^{l_{deep}}$ contained by the ${\cal C}$ cell assigned. The overall complexity is therefore $O(|{\cal C}| + |{\cal C}^{l_{deep}}|)$.

\subsection{Moving objects indexing}
\label{app: moving objects indexing complexity}

In Section \ref{sec: knn objects indexing} we introduced the \emph{moving objects indexing} phase. Here we detail the related complexity.

The overall complexity of this subphase is $O(|P| + b \cdot |P| + |P| + 2|P| + |\hat{\cal C}|)$: the first term is related to the $l_{deep}$ Morton codes parallel computation, while the second term is related to the subsequent sorting operation (by means of Radix Sort). The third term is due to the lookups in $z_{map}$ needed to retrieve the final leaves identifiers. Finally, the fourth term is due to the double scan over the set of objects needed to detect the discontinuities between objects belonging to different cells, while $|\hat{\cal C}|$ represents the amount of active cells for which we actually have to write out the related indexing information. Since $|\hat{\cal C}| \ll |P|$, the overall complexity is in the order of $O(|P|)$.

\subsection{Iterative k-NN queries computation}

In the following we detail some of the complexities characterizing the main operations carried on during the \emph{iterative k-NN queries computation} phase (Section \ref{sec: knn computation}).

\subsubsection{First iteration}

\paragraph{Query indexing and task materialization.}
\label{app: queries indexing complexity}
\vspace{1em}
This phase, which is described in Section \ref{par: knn query indexing task materialization}, has a complexity which can be derived equivalently to the one characterizing the \emph{moving objects indexing} phase, so the reader should refer to Appendix \ref{app: moving objects indexing complexity}). Consequently, the complexity of this phase is in the order of $O(|Q|)$.

\paragraph{Distance computations.}
In the following we describe the complexity of the \emph{distance computations} phase presented in Section \ref{par: knn distance computations}. Aside from specific patterns used to write out queries results, strategies based on caching or coalescing are, on the whole, almost identical. For this reason, we take Algorithm \ref{lst:knn first iteration - distance computation} as a point of reference. If we consider the operations carried on within each task, the complexity of the distance computation phase is mainly dictated by the scans over the set of objects belonging to each cell $c \in {\overline{C}}$, as well as by the k-selection algorithm. 
The overall number of scans inside Algorithm \ref{lst:knn first iteration - distance computation} are two, one at line \ref{findMinMaxDist} and the other one at lines \ref{cycleCopyObjects}-\ref{cycleCopyObjects end}; therefore, the related complexity is equal to $O(2 \cdot |\{P \cap c\}|)$.
For what is related to the k-selection algorithm we have that the number of iterations strongly depends on local densities affecting the spatial distribution inside a cell. If \emph{d} represents the minimum distance between pairs of objects in \emph{c} and \emph{numBins} represents the amount of bins used, such figure is equal to \cite{alabi2012fast}:
\begin{equation}
\label{eq: k-selection complexity}
O \left( \left\lceil \left| log_{numBins} (\frac{dist_{max} - dist_{min}}{d}) \right| \right\rceil \right),
\end{equation}
Summing up the considerations done above, if we denote the amount of iterations yielded by \emph{findKDist} in the worst case as \emph{maxIterations}, the overall complexity of the distance computation phase becomes:
\begin{equation}
\label{eq: distance complexity}
O \bigg( (2 + maxIterations) \cdot |\{P \cap c\}| \bigg).
\end{equation}

\subsubsection{Subsequent iterations}

in the following we review the complexities of the main operation carried on during any subsequent iteration (Section \ref{sec: knn subsequent iterations}).

\paragraph{Massive tree navigation.}
\vspace{1em}
Query-wise, the complexity of this subphase is dictated by the set of active queries in the direction considered, i.e., $O(|Q_{processed}|)$; if we focus on a single query, the worst case scenario corresponds to visiting all quadtree nodes when such quadtree corresponds to a uniform grid, given that we use $z_{map}$ to recover leaves identifiers. If $l_{deep}$ is the deepest quadtree level, this corresponds to a complexity equal to $O(\sum_{l=0}^{l_{deep}}4^l) = O(\frac{1-4^{l_{deep}-1}}{1-4})$.
However, these complexities tell us little on the overall amount of iterations we have to expect from our approach, since this depends on portions of the tree involved by queries, which in turn depend on a complex mix of factors such as the objects spatial distribution, how such distribution may vary over the time, the quadtree height, how objects spread across the quadtree leaves and the amount of nearest neighbours per query to compute (\emph{k}).

\paragraph{Query indexing and task materialization.}
The same reasoning done in Appendix \ref{app: queries indexing complexity} applies here as well, however considering that the set of queries indexed at any iteration is limited to those active in the direction considered.

\paragraph{Distance computations.}
The overall complexity of this subphase can be determined by means of Equations \ref{eq: k-selection complexity} and \ref{eq: distance complexity}, since the underlying algorithms are derived from the ones used in the \emph{distance computation} phase performed during the first iteration (Section \ref{par: knn distance computations}). 
More precisely, if \emph{c} denotes the cell considered for a given query \emph{q}, we have that the complexity related to the determination of the (up to) \emph{k} nearest objects in \emph{c} is, in the worst case
$$O \bigg( (2 + maxIterations) \cdot | \{p | p \in c \land d(p,q) < MAXDIST_q \} | \bigg),$$
while the operations needed to fuse the set of (up to) \emph{k} nearest objects in \emph{c} with the (up to) \emph{k} nearest objects in the query result set computed so far is, in the worst case,
$$O \bigg( (2 + maxIterations) \cdot 2k \bigg).$$

\end{document}